\documentstyle[preprint,aps]{revtex}
\begin{document}
\draft
\preprint {SNUTP-94-xx }
\title{Action-Angle Variables
for Complex \\
Projective Space  and
Semiclassical Exactness}

\author{Phillial Oh\thanks{E-mail
 address:ploh@yurim.skku.ac.kr}}
\address{Department of Physics,
Sung Kyun Kwan University,
Suwon 440-746,  KOREA}
\author{Myung-Ho Kim\thanks{E-mail
 address:mhkim@yurim.skku.ac.kr}}
\address{ Department of Mathematics,
 Sung Kyun Kwan University,
 Suwon 440-746, KOREA }

\maketitle

\begin{abstract}
We  construct the action-angle variables
of a classical integrable  model
defined on complex projective phase space
and  calculate the quantum
mechanical propagator in the coherent state path integral
representation using the stationary phase approximation.
We show  that the resulting expression
for the propagator coincides
with the exact propagator which was
obtained by solving the time-dependent Schr\"odinger equation.
\end{abstract}

\vspace{2 cm}

\noindent May, 94
\newpage

There exist several examples of physical systems in which the
semiclassical approximation to a path integral gives an exact
quantum mechanical results\cite{zinn}.
Recently, this semiclassical
exactness gained new interests\cite{ston,pick,blau,blau1,dyks}
in relation to the Duistermaat-Heckman (DH) integration formula
\cite{duis,berl,witt} which  states that the integral of the
exponential of the Hamiltonian $H$ with Liouville measure
localizes to the critical
points of  $H$ if $H$ which is defined on compact phase space
${\cal M}$ generates torus action,
and  subsequently the stationary phase
approximation gives the exact results.

The existence of torus action on the phase space
${\cal M}$ by symplectic diffeomorphism is
one of the essential ingredients of the DH formula.
In this respect, it could find some
application to completely integrable system\cite{arno}.
Let us consider  a completely integrable system on  phase space
${\cal M}$ of dimension $2N$ in which $N$ conserved quantities
 $J_m$ $(m=1,\cdots,N)$ with Hamiltonian $H=H(J_m)$ are in
 involution,
\begin{equation}
\{H, J_m\}=\{J_m,J_n\}=0,
\end{equation}
where the Poisson bracket is defined by the
symplectic structure $\Omega=
\frac{1}{2}\Omega_{AB}d\xi^A\wedge d\xi^B$ of  ${\cal M}$
\begin{equation}
 \{f,g\}=\Omega^{AB}\partial_Af\partial_Bg,\label{poi}
\end{equation}
 with $ f,g\in C^\infty({\cal M})$.

The Liouville's theorem states that if the manifold
${\cal M}_J$ defined by the level set of the functions $J_m$
\begin{equation}
{\cal M}_J=\{\xi:J_m(\xi)=j_m\}
\end{equation}
with a constant $j_m$ is compact, then
it is a smooth $N$-dimensional manifold
diffeomorphic to the torus
\begin{equation}
T^N=\{(\phi_1, \cdots, \phi_N)mod \ \ 2\pi\}.
\end{equation}
 and invariant with respect to the symplectic diffeomorphism
generated by the Hamiltonian. In this case, we can find the
action variables ${\vec I}=(I_1,\cdots,I_N)$
conjugate to the angle ${\vec\phi}=(\phi_1, \cdots, \phi_N)$
so that the original symplectic structure
is expressed by the Darboux two form
\begin{equation}
\Omega=\sum_{m=1}^NdI_m\wedge d\phi_m.\label{cano}
\end{equation}
The original Hamiltonian $H(J_m)$ is a  function of such
 $I_m^\prime$s, $H=H({\vec I})$ and  the system is  solved
completely:
\begin{equation}
{\vec I}(t)={\vec I}(0), \quad {\vec\phi}(t)={\vec\phi}(0)+
{\vec\omega}({\vec I})t
\end{equation}
where ${\vec\omega}({\vec I})=
\partial H({\vec I})/\partial {\vec I}$.

The Hamiltonian vector field $X_m$ associated with the
action variable
$I_m$ generates the m-th circular action:
\begin{equation}
X_m=\frac{\partial}{\partial \phi_m}\quad (m=1,\cdots,N)
\end{equation}
Therefore, if the Hamiltonian is a linear combination of the
action variables
\begin{equation}
H=\vec\omega\cdot\vec I+H_0\label{hamil}
\end{equation}
 with some constant $\omega_m$ and $H_0$ which are
independent of
 $\vec I$, then $H$ generates
the torus action  $T^N$ and
DH formula can be applied to the integral of
the exponential of $H$. In such a case, the stationary phase
approximation should yield the exact
quantum mechanical results.
However,  if the Hamiltonian is quadratic or higher order
functions of the
action variables, the semiclassical exactness does not hold.
For example, if $H=I_1^2$, $X_H$ generates $\phi^1$
circle action,
but in general this circle action  does not have a
constant angular
velocity on the phase space and
 DH formula does not hold\cite{duis}.

In this paper, we  present  another example of physical system
in which
the semiclassical approximation gives the exact
quantum mechanical
results using the above observation.
It is a classical integrable  model of $SU(N+1)$ isospin
defined on complex projective phase space in the
external  magnetic field  where the action-angle variables
are explicitly constructed\cite{oh4}.
We   calculate the quantum
mechanical propagator in the coherent state
path integral method
using the semiclassical approximation
and show  that the result  agree with the
exact propagator which was
obtained by solving the
time-dependent Schr\"odinger equation\cite{oh4}.
We note  that our result is the
$SU(N+1)$ generalization of the
semiclassical exactness of $SU(2)$ spin models which already
exist in the
literature\cite{ston,klau,niel,kesk}.

The explicit construction of the action-angle variables for
$CP(N)$ manifold can be made traceable
by performing the symplectic reduction
of $CP(N)$  from  $S^{2N+1}$,
$CP(N) \simeq S^{2N+1}/U(1)$\cite{lo2}.
In terms of a complex column vector $z=(z_0,z_1,\cdots,z_N)^T
\in {\bf C}^{N+1}$  and its complex conjugate
$\bar z=( z^\ast_0, z^\ast_1,\cdots , z^\ast_N)$,
$S^{2N+1}$ is defined by the constraint
\begin{equation}
\phi=\bar z z-1= 0.
\label{sphere}
\end{equation}
Let us define  the symplectic structure on ${\bf C}^{N+1}$ by
\begin{equation}
\Omega_{{\bf C}}=2iJd\bar z\wedge dz\label{symp}
\end{equation}
with a  constant $J$. Then for $f,g\in C^\infty({\bf C}^{N+1})$,
we have
\begin{equation}
\{f,g\}=-\frac{i}{2J}\sum_{I=0}^N
\left(\frac{\partial f}{\partial {\bar z}_I}
\frac{\partial g}{\partial z_I}
-\frac{\partial f}{\partial z_I}
\frac{\partial g}{\partial {\bar z}_I}\right)
\label{poisson}
\end{equation}
The next step is to introduce standard coordinates of
$CP(N)$ by
$\xi_m=z_m/z_0(z_0\neq 0, m=1,2,\cdots,N)$ and
to make coordinate
transformation from $z_I$ to $(z_0,\xi_m)$.
To make reduction  to the  $CP(N)$ manifold,
 we  choose a gauge condition such that ${\bar z}_0
 =z_0$. We note that similar
gauge condition was chosen in reduction to
 the maximal orbits of $SU(N+1)$ group\cite{alek}.
Then, the solution to the constraint Eq.(\ref{sphere})
 is given by
\begin{equation}
{\bar z}_0=z_0=\frac{1}{\sqrt{1+\vert \xi\vert^2}}
\label{sol1}
\end{equation}
where $\vert\xi\vert^2=
\sum_{m=1}^N\xi^*_m\xi_m\equiv\bar\xi\xi.$
By substituting $z_0=\frac{1}{\sqrt{1+\vert \xi\vert^2}}$ and
$z_m=\frac{\xi_m}{\sqrt{1+\vert \xi\vert^2}}$  into
Eq.(\ref{symp}), we find that  the
symplectic structure Eq.(\ref{symp})
descends to the one  which is given by the
Fubini-Study metric on  $CP(N)$
\begin{equation}
\Omega=2iJ\left[\frac{d{\bar\xi}\wedge d\xi}{1+
\vert\xi\vert^2}-\frac{( \xi d{\bar\xi})\wedge({\bar\xi}d\xi)}
{(1+\vert\xi\vert^2)^2}\right].
\label{omeg}
\end{equation}
Also, the Poisson bracket  Eq.(\ref{poisson})
descends to $CP(N)$ by
\begin{equation}
\{F,H\}=-i\sum_{m,n}g^{mn}\left(\frac{\partial F}
{\partial \xi^*_m}\frac{\partial H}{\partial \xi_n}-
\frac{\partial F}{\partial\xi_n}\frac{\partial H}
{\partial  \xi^*_m}\right),
\end{equation}
where $g^{mn}$ is the inverse of the Fubini-Study metric
given by
\begin{equation}
g^{mn}=\frac{1}{2J}(1+\vert\xi\vert^2)
(\delta_{mn}+ \xi^{\ast}_m\xi_n).\label{metric}
\end{equation}

The search for explicit expression for
the action-angle variables on $CP(N)$
is facilitated by the observation that the
Hamiltonian function
which generates the circle action of $z_m\in {\bf C}^{N+1}$
coordinate is given by the quadratic function
 $2Jz^\ast_mz_m$ (no sum on $m$) which can be easily checked
\begin{equation}
\{z_n,\epsilon 2Jz^\ast_mz_m\}=i\epsilon z_m\delta_{mn}
\end{equation}
using the Eq.(\ref{poisson}).
Hence on $CP(N)$, the Hamiltonian
function $I_m$ which generates the
circle action on $\xi_m$ plane can be expressed as
\begin{equation}
I_m=\frac{2J\xi^{\ast}_m\xi_m}{1+\vert\xi\vert^2}.
\label{action}
\end{equation}
The explicit form of the action-angle variables can be
made more palpable by the use
of stereo graphical projection.
Let us introduce the polar angles
$(\theta_1,\cdots,\theta_N) (0\leq\theta\leq \pi)
$ and $(\phi_1, \cdots, \phi_N)$ via
\begin{eqnarray}
\xi_1 & = & \tan (\theta_1/2)\cos(\theta_2/2)e^{-i\phi_1}
\nonumber\\
\xi_2 & = & \tan (\theta_1/2)\sin(\theta_2/2)\cos(\theta_3/2)
e^{-i\phi_2}\nonumber\\
     & \cdot & \cdots\nonumber\\
\xi_{N-1} & = & \tan (\theta_1/2)\sin(\theta_2/2)\cdots
\sin(\theta_{N-1}/2)\cos(\theta_N/2)e^{-i\phi_{N-1}}
\nonumber\\
\xi_{N} & = & \tan (\theta_1/2)\sin(\theta_2/2)\cdots
\sin(\theta_{N-1}/2)\sin(\theta_N/2)
e^{-i\phi_{N}}\label{polar}
\end{eqnarray}
Using Eq.(\ref{action}), we find the following expression for
the action variables  $I_m$
\begin{eqnarray}
I_m & = &2J\sin^2(\theta_1/2)  \cdots \sin^2(\theta_m/2)
\cos^2(\theta_{m+1}/2) \quad (m<N)\nonumber\\
I_N & = &2J\sin^2(\theta_1/2)\cdots
\sin^2(\theta_{N-1}/2)\sin^2(\theta_N/2)
\end{eqnarray}
One can  check  that substitution of Eq.(\ref{polar}) into
Eq.(\ref{omeg}) produces Eq.(\ref{cano}) with $I_m$
given by the above formula. So our model Hamiltonian
is given by a linear combination of action variables
given by Eq.(\ref{hamil}) and (\ref{action}).
The Hamiltonian vector field associated to it generates
$T^N$ torus action given by
\begin{equation}
X_H=i\sum_{m=1}^N\omega_m(\xi_m\frac{\partial}
{\partial\xi_m}-\xi^\ast_m\frac{\partial}
{\partial \xi^\ast_m})
=\sum_{m=1}^N\omega_m\frac{\partial}{\partial \phi_m}.
\end{equation}
The classical solution describes a
{\it conditionally-periodic motion}\cite{arno}:
\begin{equation}
\vec\theta(t)=\vec\theta(0),
 \quad \vec\phi(t)=\vec\phi(0)+\vec\omega t.
\label{solx}
\end{equation}

We perform the path integral of our model using
the coherent state  method\cite{klau1,pere,zhan}.
 Let us consider
$\vert 0\rangle$, the highest weight state annihilated by all
positive roots of $SU(N+1)$ algebra in Cartan basis.
Then for $CP(N)$ with given $P
\equiv 2J$ $(P\in {\bf Z}^+)$ we have an irreducible
representation  $(P,0,\cdots,0)$ of $SU(N+1)$ group\cite{kiri}
and there are precisely  $N$ negative roots
$E_\alpha, \alpha=1,2,\cdots,N$
 such that $E_\alpha\vert 0\rangle\neq\vert 0\rangle.$
 Let us label $\{E_\alpha\}=\{E_m\}$.
We define a coherent state on $CP(N)$ corresponding
to the point $\xi=(\xi_1,\cdots,\xi_N)$ by \cite{pere,zhan,gitm}
\begin{equation}
\vert P,\xi\rangle=\exp(\sum_m\xi_mE_m)\vert 0\rangle
\end{equation}
Notice that this definition differs from the usual one by
the normalization factor.
We have chosen this definition here because in the subsequent
analysis, $\bar\xi$ and $\xi$
can be treated independently and the overspecification problem
can be side-stepped\cite{klau,fadd2,brow}. We denote
$\vert P,\xi\rangle=\vert\xi\rangle$ from now on.
The coherent states which we have defined on
 $CP(N)$ have the
following two  properties which are essential
in the path integral formulation. One is the resolution
of unity,
\begin{equation}
\int D\mu({\bar \xi},\xi)\frac{ \vert
\xi\rangle \langle \xi\vert}{ (1+\vert\xi\vert^2)^{2J}}=I,
\label{id}
\end{equation}
where $D\mu({\bar \xi},\xi)=c d\bar\xi d\xi/
(1+\vert\xi\vert^2)^{N+1}$ with a constant $c$  is the
Liouville measure\cite{zhan}.
The other is reproducing kernel,
\begin{equation}
\langle \xi^{\prime\prime}\vert \xi^\prime\rangle=
(1+\bar\xi^{\prime\prime}\xi^\prime)^{2J}.\label{ker}
\end{equation}

We are interested in evaluating the propagator
\begin{equation}
G(\bar\xi^{\prime\prime},\xi^\prime;T)
= \langle \xi^{\prime\prime}\vert e^{-i{\hat H}T}\vert
\xi^\prime\rangle.
\end{equation}
Let us divide the time interval $T=t^{\prime\prime}-t^{\prime}$
by $P+1$ steps with $\bar\xi(P+1)=\bar\xi^{\prime\prime}$
and $\xi(0)=\xi^\prime$,
and let $\epsilon=T/(P+1)$.
 Inserting Eq.(\ref{id})  repeatedly
\begin{equation}
G(\bar\xi^{\prime\prime},\xi^\prime;T)=
\lim_{\epsilon\rightarrow 0}\int\cdots\int
\prod_{p=1}^{P}\frac{D\mu(p)}{(1+\vert\xi(p)\vert^2)^{2J}}
\prod_{p=1}^{P+1}\langle\xi(p)\vert\xi(p-1)\rangle
\left(1-i\epsilon\frac{\langle\xi(p)
\vert\hat H\vert\xi(p-1)\rangle}
{\langle\xi(p)\vert\xi(p-1)\rangle}\right)
\end{equation}
 and using Eq.(\ref{ker}), we have
\begin{eqnarray}
\prod_{p=1}^{P+1} <\xi(p)\vert \xi(p-1)>
& \approx &
\prod_{p=1}^{P+1}(1+\vert\xi(p)\vert^2)^{2J}
\left(1-\frac{2J\bar\xi(p)d\xi(p)}
{1+\vert\xi(p)\vert^2}\right)\nonumber\\
& = &\prod_{p=1}^{P+1}(1+\vert\xi(p)\vert^2)^{2J}
\exp\left\{\epsilon\left(\frac{-2J\bar\xi(p)d\xi(p)/\epsilon}
{1+\vert\xi(p)\vert^2}\right)\right\}
\end{eqnarray}
Hence in the limit $\epsilon\rightarrow 0$
we obtain the following expression
\begin{equation}
G(\bar\xi^{\prime\prime},\xi^\prime;T)=
\int D\mu\exp\left\{2J\log(1+\bar\xi^{\prime\prime}
\xi(t^{\prime\prime}))+
i\int_{t^\prime}^{t^{\prime\prime}}dt\left[i
\frac{2J{\bar \xi}\dot \xi}{1+\vert\xi\vert^2}
-  H(\bar\xi, \xi)\right]\right\}\label{prop}
\end{equation}
where
$ H(\bar\xi, \xi)=\langle \xi\vert {\hat H}(\bar\xi, \xi)\vert
\xi\rangle/\langle \xi\vert \xi\rangle$ is the classical
Hamiltonian  given by the Eq.(\ref{hamil}) and (\ref{action}).
The boundary conditions in the path integral
is given by $\xi(t^\prime)=\xi^\prime$ and
$\bar\xi(t^{\prime\prime})=\bar\xi^{\prime\prime}$. Also
$G(\bar\xi^{\prime\prime},\xi^\prime;T)
{\Big\vert}_{T\rightarrow 0}=
(1+\bar\xi^{\prime\prime}\xi^\prime)^{2J}$.
We introduced $ \xi(t^{\prime\prime})$
which is only a  superfluous variable
because the result of path integral
 Eq.(\ref{prop}) does not depend on this variable.
It depends only on
$\bar\xi^{\prime\prime}$ and $\xi^\prime$.
 The equations of motion are
\begin{equation}
i\dot\xi_m=g^{\ast mn}\frac{\partial H(\bar\xi,\xi)}
{\partial \xi^{\ast}_n},\quad  i\dot\xi^{\ast}_m=-g^{mn}
\frac{\partial  H(\bar\xi,\xi)}{\partial \xi_n}.
\end{equation}
Using the Hamiltonian given in Eq.(\ref{hamil})
and (\ref{action}), we get the following
\begin{equation}
 \dot\xi^{\ast}_m-i\omega_m\xi^{\ast}_m=0,
\quad \dot\xi_m+i\omega_m\xi_m=0,\label{equation}
\end{equation}
where no summation on the index $m$ is assumed. We note that
our model is equivalent to a
collection of $N$ harmonic oscillator in coherent
state representation \cite{fadd2,brow}
although the Hamiltonian appears to be highly nonlinear.
The solutions are given by
\begin{equation}
\xi^{\ast}_m(t)=\xi^{\ast\prime\prime}_me^{i\omega_m
(t-t^{\prime\prime})}\quad
\xi_m(t)=\xi^{\prime}_me^{-i\omega_m(t-t^\prime)}
\label{soll}
\end{equation}

Now, we evaluate the propagator Eq.(\ref{prop}) around the
classical solution using stationary phase method\cite{klei}.
 Denoting the above classical solution by $\bar\xi_{cl}$ and
$\xi_{cl}$ and expanding around the classical solutions
\begin{equation}
\bar\xi=\bar\xi_{cl}+\delta\bar\xi,
\quad \xi=\xi_{cl}+\delta\xi
\end{equation}
with boundary conditions $\delta\bar\xi(t^{\prime\prime})=
\delta\xi(t^{\prime})=0$,
we find the following expression for the propagator:
\begin{equation}
G(\bar\xi^{\prime\prime},\xi^\prime;T)=K(T)(1+\bar\xi^{\prime\prime}_{cl}\xi_{cl}
(t^{\prime\prime}))^{2J}\exp(iS(\bar\xi_{cl}, \xi_{cl},T)
\label{result}
\end{equation}
Here $K(T)$ is the Van Vleck-Pauli-Morette
determinant\cite{klei}
coming from the Gaussian integration of the
fluctuations $\delta\bar\xi$
and $\delta\xi$.
Its explicit form is given by $K=C_0/\sqrt{det M}$
(with $C_0$ a constant) where
\begin{equation}
M=\left(\begin{array}{cc}
-\delta^{mn}\frac{d}{dt}-i\partial^{m}
g^{\ast nl}\partial^{\ast l}H &
-i\partial^{\ast m}g^{\ast nl}\partial^{\ast l}H \\
-i\partial^{m}g^{nl}\partial^l H &
\delta^{mn}\frac{d}{dt}-
i\partial^{\ast m}g^{nl}\partial^{ l}H
\end{array}\right).
\end{equation}
Using the classical equations of motion Eq.(\ref{equation}),
 we have
\begin{equation}
M=\left(\begin{array}{cc}
-\delta^{mn}(\frac{d}{dt}+i\omega_m) & 0 \\
0 & \delta^{mn}(\frac{d}{dt}-i\omega_m)
\end{array}\right).
\end{equation}
{}From the above equation,
we see that $det M$ is equal to $2N$ product
of harmonic oscillator factor
($N$ harmonic oscillator with frequency $\omega_m$ and the other
 $N$ with $-\omega_m$)
and  a simple lattice calculation
shows that each harmonic oscillator factor
is equal to one. So we have  $K(T)=C_0$.
Substituting the classical solution Eq.(\ref{soll})
 and $K(T)=C_0$ into Eq.(\ref{result}),  we obtain
\begin{equation}
G(\bar\xi^{\prime\prime},\xi^\prime;T)=C_0\left( 1+
\sum_{m=1}^N\xi^{\ast\prime\prime}_m\xi^{\prime}_m
\exp(-i\omega_mT)\right)
^{2J}\exp(iH_0T).
\end{equation}
The above expression is equal to the exact propagator obtained
by solving the time-dependent schr\"odinger equation\cite{oh4}
(with $C_0=1$ which is fixed by
the boundary condition $G(\bar\xi^{\prime\prime},\xi^\prime;T)
{\Big\vert}_{T\rightarrow 0}=
(1+\bar\xi^{\prime\prime}\xi^\prime)^{2J}$).

In summary, we  constructed the action-angle variables
of the completely integrable  model
defined on $CP(N)$ manifold
by using the symplectic reduction
of $CP(N)$  from  $S^{2N+1}$.
Then, adopting the coherent state path integral method,
we showed that the stationary phase approximation
of the integrable model yields
the exact quantum mechanical propagator, thus
providing the  $SU(N+1)$ generalization of the
semiclassical exactness of $SU(2)$ spin models
\cite{ston,klau,niel,kesk}.
 Extension to other coadjoint orbits of Lie group
is also conceivable and will be reported elsewhere.

\acknowledgments
This work is supported by Ministry of Education through
the Research Institute of Basic Science and in part by
the KOSEF through C.T.P. at S.N.U.

\end{document}